\documentclass{article}
\usepackage[T1]{fontenc}
\usepackage{fourier}
\usepackage[english]{babel}															
\usepackage[protrusion=true,expansion=true]{microtype}	
\usepackage{amsmath,amsfonts,amsthm, amssymb} 
\usepackage[pdftex]{graphicx}	
\usepackage{url}
\usepackage{physics}
\usepackage[square,sort,comma,numbers]{natbib}
\usepackage{hyperref}
\hypersetup{colorlinks=true, linkcolor=blue, citecolor=violet}
\usepackage{xcolor}
\usepackage{tensor}
\usepackage{tikz}
	\usetikzlibrary{knots}
	\usetikzlibrary{calc}
	\usetikzlibrary{positioning}
	\usetikzlibrary{angles,quotes}

\usepackage{fancyhdr}
\numberwithin{equation}{section}		
\numberwithin{figure}{section}			
\numberwithin{table}{section}				

\author{
 Samuele Faglioni\\
  \small Int. Institute for Sustainability with Knotted Chiral Meta Matter ($\text{WPI-SKCM}^2$)\\
  \small Hiroshima University\\
  \small 1-3-1 Kagamiyama, Higashi-Hiroshima, Hiroshima 739-8526, Japan \\
  \small \texttt{sfaglioni@hiroshima-u.ac.jp} \\
}

\title{Tight composites knots and chirality}

\definecolor{green_slide_texts}{HTML}{006400}

\begin{document}
\maketitle

\begin{abstract}
Chirality is known to play a central role in the properties of many physical systems across a wide range of spatial and temporal scales. Chemical and optical properties of materials are only two of the many examples where transformation properties under reflection symmetry become relevant in describing a real-world system: within this context, the word \textit{enantiomers} is used to describe two different types of geometric shapes related by a reflection, called \textit{left-handed} or \textit{right-handed} enantiomers, in reference to the definition of chirality and handedness of screws presented by Maxwell \cite{Maxwell1873} in its treatise. In this short communication, the relation between chirality and the geometric shape of tight composite knots is discussed using arguments from the linear elastic theory of ropes. The results presented here serve as the starting point for a more general analysis, which we intend to pursue in future investigations.
\end{abstract}

\section{Introduction}
The elastic theory of ropes provides many insights about the relation between geometry and topology of closed loops. While knot theory offers useful tools to classify knot types, these are not sufficient to explain all the physical properties of real knots, as such properties can be changed using topological invariant moves. To address this, methods from linear elastic theory help to link the geometry of a closed elastic loop with external forces. 

Chirality constitutes a fundamental property in various physical systems, although its manifestations may differ significantly depending on the context. For example, in the case of magnetic materials, chiral properties determine the chemical structure and play a fundamental role in the interaction with an external magnetic field \cite{Inoue2021}. In the case relevant to this study, chirality also affects the mechanical properties of elastic knots. A common example is the difference in stiffness of the Square knot and the Granny knot, obtained starting from the trefoil knot. The result of a reflection transformation on the trefoil knot, denoted as $3_1$ in the Alexander-Briggs notation, is called the \textit{mirror image} of $3_1$ and is denoted with $3_1^*$. Then, the trefoil knot is said to be \textit{chiral} as there are no continuous transformations able to deform $3_1^*$ into $3_1$. Conversely, a knot $K$ that is topologically equivalent to its mirror image is said to be \textit{achiral}\footnote{The simplest example of an achiral knot is $4_1$.}. From the trefoil knot, it is possible to construct two distinct topological configurations using the direct sum, with different chirality. The first configuration corresponds to the chiral composite $3_1\# 3_1$, commonly known as the Granny knot. The latter corresponds to the achiral composite $3_1\# 3_1^*$, commonly known as the Square knot. In the context of chemistry, the latter is known as \textit{racemic pair}. 

In this brief report, we examine the relationship between chirality and tight composite knots using arguments based on linear elastic theory of ropes \citep{Audoly2010}. Specifically, we calculated the curvature and torsion contributions to the elastic energy for the chiral configuration of tight composite knots composed of two elements. As a result, it was found that, in the tight configuration, the composite made by $3_1\# 3_1^*$ has higher value of the elastic energy than $3_1\# 3_1$ when the bending and torsion stiffness are normalized to unit. This result agrees with the difference between the Square and the Granny knot, providing an example of two different shoelace tying methods. 

A generalization of the result from composites of the trefoil is presented among the class of chiral prime composite knots up to six crossings. Using the knot tightening software \texttt{ridgerunner} \citep{Cantarella2010}, we obtain a precise geometrical configuration of composite knots which also has an important topological interpretation \citep{Stasiak1998}. The analysis was done on a small set of composite configuration and has to be intended as the starting point for a general mathematical fashion, which is our proposal for ongoing work. 

\section{Methods}
The analysis of the tight configuration of composite knots has advantages in terms of geometry and topology. Among all the possible geometries allowed for closed loops with fixed topology, the definition of the ideal shape of a knot type plays an important role as a bridge between its geometrical and topological information. However, this definition is not unique and depends on a criterion. The one provided by Katritch et al. \cite{Katritch1996} has an intuitive meaning that can be applied to the standard notion of elastic knots. Following their discussion, the ideal shape of a knot type is the ``shortest piece of tube that can be closed to form the knot''. The mathematical function able to detect the ideal shape is the \textit{thickness} \citep{Gonzales1999}, and its reciprocal provides a quantity called ropelength, which offers a topological classification of knot types \citep{Buck1999}. 

\begin{figure}[h!]
		\centering
		\begin{tikzpicture}
			\draw[-] (-1, 0)--(0, 0.5)--(1, 0);
			\draw[-] (-0.3, 0.35) to[bend right=45] (0.3, 0.35);
			\draw[dashed, color=gray] (-1, 0) -- (-2, -0.25);
			\draw[dashed, color=gray] (1, 0) -- (2, -0.75);
			\node[circle,fill=orange,inner sep=0pt,minimum size=4pt] at (-1,0) {};
			\node[circle,fill=orange,inner sep=0pt,minimum size=4pt, label=above:{$\textcolor{orange}{i}$}] at (0,0.5) {};
			\node[circle,fill=orange,inner sep=0pt,minimum size=4pt] at (1,0) {};
			\node at (0, 0) {$\theta_i$};
			\node at (0,-0.8) {$\bar{\theta}_i = \theta_i \frac{L(e_i, e_{i+1})}{2}$};
			\node at (-0.75, 0.45) {$e_i$};
			\node at (0.75, 0.45) {$e_{i+1}$};
 		\end{tikzpicture}
 		\hspace{1cm}
 		\begin{tikzpicture}
			\draw[-] (-1, 0)--(0, 0.5)--(1, 0)--(2, 0.25);
			\draw[dashed, color=gray] (-1, 0) -- (-2, -0.25);
			\draw[dashed, color=gray] (2, 0.25) -- (3, 0.45);
			\node[circle,fill=orange,inner sep=0pt,minimum size=4pt, label=below:{$\textcolor{orange}{A}$}] at (-1,0) {};
			\node[circle,fill=orange,inner sep=0pt,minimum size=4pt, label=above:{$\textcolor{orange}{B}$}] at (0,0.5) {};
			\node[circle,fill=orange,inner sep=0pt,minimum size=4pt, label=below:{$\textcolor{orange}{i}$}] at (1,0) {};
			\node[circle,fill=orange,inner sep=0pt,minimum size=4pt, label=above:{$\textcolor{orange}{C}$}] at (2,0.25) {};
			\node at (-0.75, 0.45) {$e_{i-1}$};
			\node at (0.65, 0.35) {$e_i$};
			\node at (1.75, -0.1) {$e_{i+1}$};
			\node at (0.5,-0.8) {$\bar{\phi}_i = \phi_i \frac{L(e_{i-1}, e_i, e_{i+1})}{3}$};
 		\end{tikzpicture}
 		\caption{Weighted curvature and torsion angle are denoted with $\bar{\theta}_i$, $\bar{\phi}_i$. The angle $\phi_i$ is the angle between the plane containing $(A, B, i)$ and the plane containing $(B, i, C)$. $L(e_i, e_{i+1})$ is the sum of the lengths of the edges $e_i$ and $e_{i+1}$.}
 		\label{fig:1}
	\end{figure}
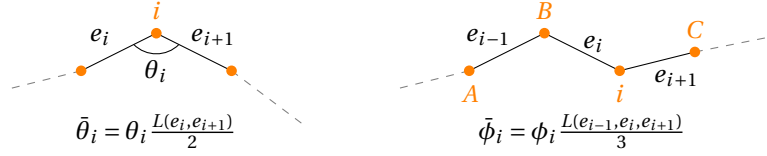

The knot tightening software \texttt{ridgerunner} provides geometric quantities of closed curves in their ideal shape, such as curvature and torsion angles. The algorithm also computes the measure of each segment composing the discrete knot. Let $L_i$ be the length of the $i$-th edge. The curvature and torsion angle $\theta_i$ and $\phi_i$ related to the $i$-th segment are computed for discrete composite knots in their ideal shape. Figure \ref{fig:1} shows the curvature and torsion angle weighted with the edge's lengths information, denoted with $\bar{\theta}_i$ and $\bar{\phi}_i$.
Following the work of \cite{Bergou2008}, the discrete Frenet curvature and torsion can be written as
\begin{equation}
 \kappa_i = \frac{1}{2}\tan(\frac{\bar{\theta}_i}{2}), \qquad  \tau_i = \frac{1}{2}\tan(\frac{\bar{\phi}_i}{2}).
\end{equation}
Using $\kappa_i$ and $\tau_i$ one can define the following quantity
\begin{equation}
 \label{eqn_1}
 U = \sum_{i} (\kappa_i^2 + \tau_i^2),
\end{equation}
where the sum is done along the closed discrete knot, which corresponds to the discrete elastic energy for a rod \citep{Audoly2010} where the bending and torsional stiffness are equal to 1. 

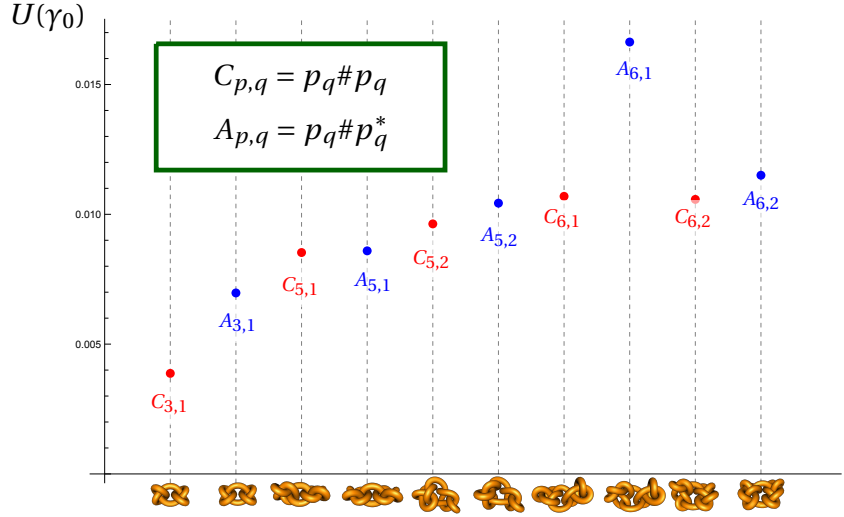
\begin{figure}[h!]
	\centering
	\scalebox{1.2}{
	\begin{tikzpicture}
		\node at (0,0) {\includegraphics[scale=0.5]{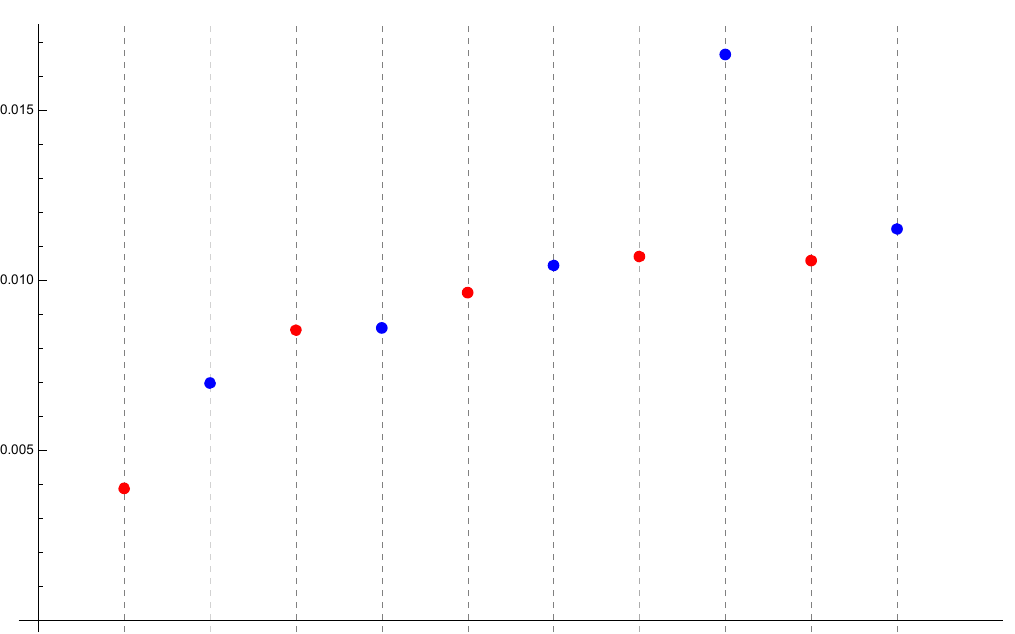}};
		
		\node at (-3.25, -2.8) {\includegraphics[scale=0.05]{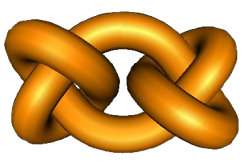}};
		\node at (-2.5, -2.8) {\includegraphics[scale=0.05]{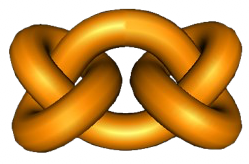}};
		
		\node at (-1.8, -2.8) {\includegraphics[scale=0.035]{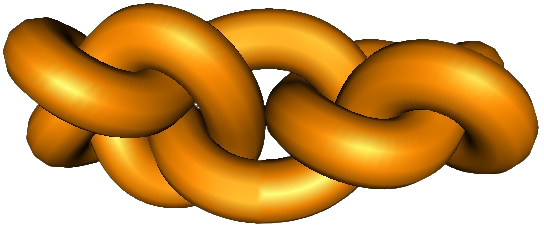}};
		\node at (-1, -2.8) {\includegraphics[scale=0.035]{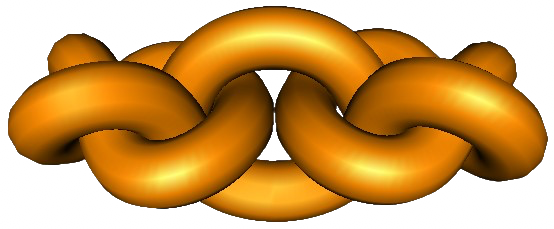}};

		\node at (-0.3, -2.8) {\includegraphics[scale=0.032]{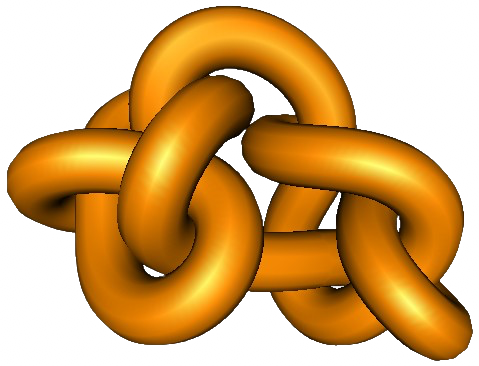}};
		\node at (0.4, -2.8) {\includegraphics[scale=0.032]{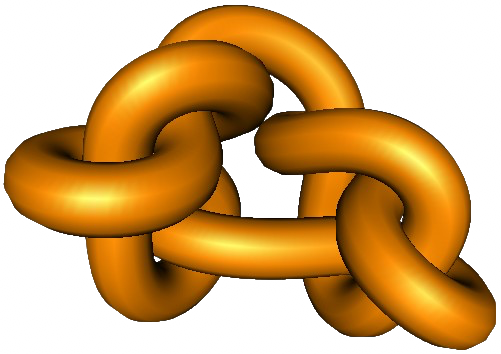}};
		
		\node at (1.1, -2.8) {\includegraphics[scale=0.03]{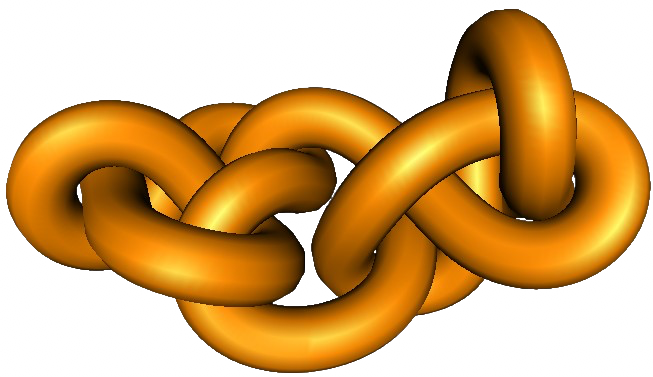}};
		\node at (1.9, -2.8) {\includegraphics[scale=0.03]{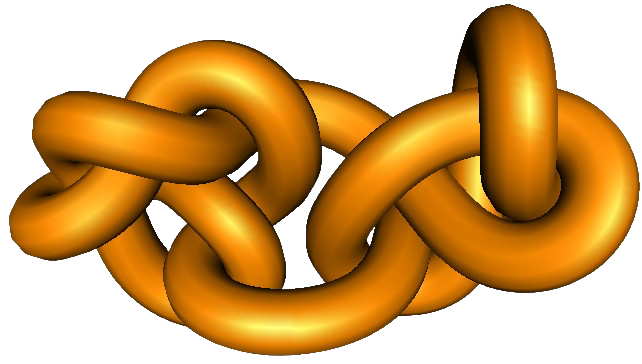}};
		
		\node at (2.55, -2.8) {\includegraphics[scale=0.032]{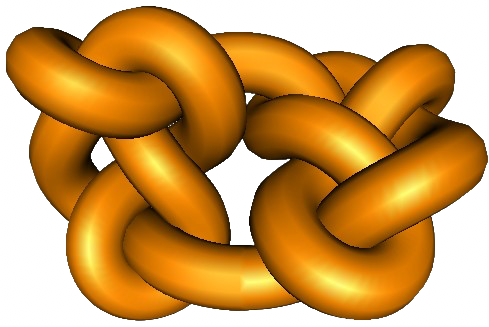}};
		\node at (3.3, -2.8) {\includegraphics[scale=0.032]{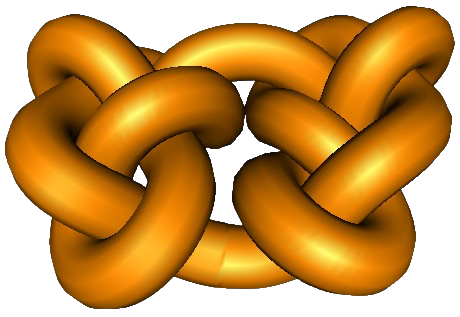}};
		
%
%
%
%
		
		\node[fill=white, text opacity=1,fill opacity=0.6] at (-3.25, -1.8) {\scriptsize $\textcolor{red}{C_{3,1}}$};
		\node[fill=white, text opacity=1,fill opacity=0.6] at (-2.5, -0.9) {\scriptsize $\textcolor{blue}{A_{3,1}}$};
		
		\node[fill=white, text opacity=1,fill opacity=0.6] at (-1.8, -0.5) {\scriptsize $\textcolor{red}{C_{5,1}}$};
		\node[fill=white, text opacity=1,fill opacity=0.6] at (-1, -0.45) {\scriptsize $\textcolor{blue}{A_{5,1}}$};
		
		\node[fill=white, text opacity=1,fill opacity=0.6] at (-0.35, -0.2) {\scriptsize $\textcolor{red}{C_{5,2}}$};
		\node[fill=white, text opacity=1,fill opacity=0.6] at (0.4, 0.05) {\scriptsize $\textcolor{blue}{A_{5,2}}$};
		
		\node[fill=white, text opacity=1,fill opacity=0.6] at (1.1, 0.25) {\scriptsize $\textcolor{red}{C_{6,1}}$};
		\node[fill=white, text opacity=1,fill opacity=0.6] at (1.9, 1.9) {\scriptsize $\textcolor{blue}{A_{6,1}}$};
		
		\node[fill=white, text opacity=1,fill opacity=0.6] at (2.55, 0.25) {\scriptsize $\textcolor{red}{C_{6,2}}$};
		\node[fill=white, text opacity=1,fill opacity=0.6] at (3.3, 0.45) {\scriptsize $\textcolor{blue}{A_{6,2}}$};
		
		\draw[-, color=green_slide_texts, fill=white, line width=1.5pt] (-3.4, 2.2) -- (-0.2, 2.2) -- (-0.2, 0.8) -- (-3.4, 0.8) -- (-3.4, 2.2);
		\node at (-1.8, 1.8) {$C_{p,q} = p_q\# p_q$};
		\node at (-1.8, 1.2) {$A_{p,q} = p_q\# p_q^*$};
		
		\node at (-4.6, 2.5) {$U(\gamma_0)$};
	\end{tikzpicture}}
	\caption{Classification of chiral and achiral composites. Although it is not possible to predict a general behavior, the quantity $U(\gamma_0)$ defined in \eqref{eqn_1} is able to distinguish between the chiral and achiral configuration of tight composites made by the same knot type.}
	\label{fig:2}
\end{figure}

The quantity \eqref{eqn_1} can be applied to the tight configuration of the composites analyzed using \texttt{ridgerunner}. The data refer to composites of prime knots up to six crossings, constructed by the connected sum of two elements of the same knot type that differ from their mirror images. Let $E$ be the set of tight knot elements and $E^*$ the set of mirror images
\begin{equation}
 E = \{3_1, 5_1, 5_2, 6_1, 6_2\}, \quad E^* = \{3_1^*, 5_1^*, 5_2^*, 6_1^*, 6_2^*\}, 
\end{equation}
Prime knots are denoted using the Alexander–Briggs notation. The analyzed configurations of tight composite knots fall into two sets:
\begin{itemize}
 \item[(i)] Chiral composites: $C = \{3_1\# 3_1, 5_1\# 5_1, 5_2\# 5_2, 6_1\# 6_1, 6_2\# 6_2\}$.
 \item[(ii)] Achiral composites: $A = \{3_1\# 3_1^*, 5_1\# 5_1^*, 5_2\# 5_2^*, 6_1\# 6_1^*, 6_2\# 6_2^*\}$.
\end{itemize}
Using \eqref{eqn_1} on the sets $C$ and $A$ defined above, one can obtain the results displayed in Figure \ref{fig:2}. According to this analysis, the quantity \eqref{eqn_1} is able to distinguish between chiral and achiral configurations of tight composite knots in the sets $C$ and $A$. Numerical values are displayed in Table \ref{tab:1} in the \nameref{Appendix}.

\section{Further directions}
Knot composites and connected sum provide a way to analyze chirality and its implications in terms of the linear elasticity of physical ropes. Recent results presented by Patil and collaborators \cite{Patil2020} discussed the mechanical stability of open 2-tangles with different configurations employing tools from knot topology and Kirchhoff elasticity. We propose that the current analysis may contribute to understanding the geometrical constraints imposed by the presence or absence of chiral configurations for composite knots in their tight structure. A key question is the role of torsion and how the connected sum behaves under different chiral constraints. As anticipated at the beginning of this short communication, a more general mathematical framework is needed to fully develop these results, a goal we intend to pursue in future work.

\section*{Acknowledgements}
I am deeply grateful to Professor Eric J. Rawdon for providing the geometric data generated with \texttt{ridgerunner}, which were essential for the analysis of composites. I also wish to thank my supervisor, Professor Katsuya Inoue, for his useful comments during the preparation of this work and for suggesting the paper written by Patil and collaborators \cite{Patil2020}, which served as the initial inspiration for the analysis discussed in this short communication.

	\nocite{*}
	\bibliographystyle{mystyle}
	\bibliography{biblio.bib}
\newpage
\section*{Appendix}\label{Appendix}
\begin{table}[h!]
\centering
	\begin{tabular}{|c|c|c|c|c|}
	\hline
	Knot type & $\sum_i \kappa_i^2$ & $\sum_i \tau_i^2$ & $U$ & Ropelength	\\
	\hline\hline
	$3_1+3_1$ & $6.00\times 10^{-7}$ & $2.00\times 10^{-5}$ & $2.00\times 10^{-5}$ & 32.74300	\\
	$C_{3,1}$ & $3.8\times 10^{-4}$ & $3.50\times 10^{-3}$ & $3.87\times 10^{-3}$ & 28.52720	\\
	$A_{3,1}$ & $3.6\times 10^{-4}$ & $6.61\times 10^{-3}$ & $6.97\times 10^{-3}$ & 28.54160	\\
	\hline\hline
	$5_1+5_1$ & $6.2\times 10^{-4}$ & $3.84\times 10^{-3}$ & $4.46\times 10^{-3}$ & 47.21460	\\
	$C_{5,1}$ & $5.8\times 10^{-4}$& $7.95\times 10^{-3}$ & $8.53\times 10^{-3}$ & 43.02610		\\
	$A_{5,1}$ & $5.8\times 10^{-4}$ & $8.01\times 10^{-3}$ & $8.59\times 10^{-3}$ & 43.05230	\\
	\hline\hline
	$5_2+5_2$ & $6.3\times 10^{-4}$ & $9.50\times 10^{-3}$ & $1.01\times 10^{-2}$ & 49.48140	\\
	$C_{5,2}$ & $5.6\times 10^{-4}$ & $9.07\times 10^{-3}$ & $9.63\times 10^{-3}$ & 43.76070	\\
	$A_{5,2}$ & $5.7\times 10^{-4}$ & $9.86\times 10^{-3}$ & $1.04\times 10^{-2}$ & 43.80490	\\
	\hline\hline
	$6_1+6_1$ & $6.2\times 10^{-4}$ & $7.72 \times 10^{-3}$ & $8.34\times 10^{-3}$ & 56.71680	\\
	$C_{6,1}$ & $6.8\times 10^{-4}$ & $1.00\times 10^{-2}$ & $1.07\times 10^{-2}$ & 51.21850	\\
	$A_{6,1}$ & $9.5\times 10^{-4}$ & $1.57\times 10^{-2}$ & $1.66\times 10^{-2}$ & 52.57750	\\
	\hline\hline	
	$6_2+6_2$ & $7.3\times 10^{-3}$ & $9.36\times 10^{-3}$ & $1.01\times 10^{-2}$ & 57.03680	\\
	$C_{6,2}$ & $6.6\times 10^{-4}$ & $9.91\times 10^{-3}$ & $1.06\times 10^{-2}$ & 51.58080	\\
	$A_{6,2}$ & $6.9\times 10^{-4}$ & $1.08\times 10^{-2}$ & $1.15\times 10^{-2}$ & 51.86390	\\
	\hline	
	\end{tabular}
	\caption{This table shows the role of connected sum in the distribution of geometric quantities. Rows $p_q + p_q$ denote the sum of two disjoint knots $p_q$. In the tight configuration, the torsion contribution exceeds the one of curvature for all the analyzed composites. Ropelength is used as an indicator to classify the topology of knot types according to \cite{Cantarella2012}.}
	\label{tab:1}
\end{table}

\end{document}